\documentstyle[aps]{revtex}
\input{epsf.tex}
\begin{document}
\draft
\title{Hydrodynamic modes in dense trapped ultracold gases}
\author{R. Combescot and X. Leyronas}
\address{Laboratoire de Physique Statistique,
 Ecole Normale Sup\'erieure*,
24 rue Lhomond, 75231 Paris Cedex 05, France}
\date{Received \today}
\maketitle

\begin{abstract}
We consider the hydrodynamic modes for dense trapped ultracold 
gases, where the interparticle distance is comparable to the scattering 
length. We show that the experimental determination of the 
hydrodynamic mode frequencies allows to obtain quite directly the 
equation of state of a dense gas. As an example we investigate the case 
of two equal fermionic populations in different hyperfine states with 
attractive interaction and in particular the vicinity of the collapse.
\end{abstract}

Most of the fascinating recent work on ultracold gases \cite{dgps} has 
been dealing with dilute situations. Naturally even in this regime 
interactions play an important role, as in the case of BEC where they 
strongly increase the size of the condensate compared to the free boson 
case. In this dilute regime the scattering length is small compared to the 
interparticle distance. However it is of great interest to explore the dense 
gas regime where scattering length and interparticle distance are 
comparable. This would lead to physical systems which are very simple 
examples of strongly interacting systems. These have much more 
complicated counterparts in condensed matter physics, such as liquid $ 
^{4}$He or $ ^{3}$He, or the electron gas in metals. This regime is 
also of major experimental interest in the search for a BCS superfluid in 
fermion gases \cite{stoofal}, since this is the range where the higher 
critical temperatures \cite{rc,holl} will be found, which should make 
the transition more accessible. Experimentally this dense regime 
corresponds to large scattering lengths, which can be reached in the 
vicinity of Feshbach resonances, as it has already been seen in optical 
traps \cite{fesh}.

\vspace{4mm}
In this paper we show that the experimental determination of the mode 
frequencies in the hydrodynamic regime allows to obtain quite 
efficiently and directly the equation of state of a dense gas. 
Hydrodynamic equations are valid in the limit of low frequency and 
long wavelength, and they have already been used to study the dilute 
Bose gas \cite{dgps,str} (with very good agreement with experiment) 
as well as the free Fermi gas \cite{brcl}. In these cases the equation of 
state is known. We show that, rather surprisingly, the analysis of the 
equation giving the mode frequencies is not much more complicated 
when the equation of state is unknown and that one can conveniently 
invert the problem and obtain the equation of state from the mode 
spectrum. Although one might consider the extension to higher 
temperature we work in the low temperature range where thermal 
effects are small and we neglect dissipation, so we deal with a perfect 
fluid. This should be valid for a strongly degenerate Fermi gas in its 
normal state. Naturally our results apply also to a superfluid when 
normal liquid effects are negligible, such as low temperature Bose 
condensates or low temperature simple scalar BCS superfluids. We 
consider for simplicity an isotropic trapping potential $V(r)$ (mostly the 
harmonic case) but, somewhat surprinsingly most of our procedures 
can be generalized to the case of anisotropic harmonic traps. Also we 
treat the 3D case, but lower dimensions can be handled in the same 
way. As an example we apply our treatment to the case of two equal 
fermionic populations in different hyperfine states with attractive 
interaction, in particular we investigate the vicinity of the collapse, a 
very interesting physical situation analogous to $ ^{7}$Li  BEC 
collapse.

\vspace{4mm}
With our hypotheses hydrodynamics reduces to Euler equation $m \,n 
\,d {\bf  v}/dt = - {\bf  \nabla}P - n {\bf  \nabla}V(r)$ supplemented by 
particle conservation $ \partial n / \partial t + {\bf  \nabla} (n {\bf  v}) = 
0 $ for density $ n ({\bf  r},t)$. The equilibrium particle density $ 
n_{0}(r) $ satisfies $ \mu (n_{0}(r)) + V(r) = \tilde{\mu }$ where  
$\tilde{\mu }$ is the overall chemical potential. Below we refer for 
short to $\mu (n )$ as the equation of state. Linearizing these equations 
around equilibrium one finds that the density fluctuation $ n_{1}({\bf  
r},t) = n ({\bf  r},t) - n_{0}(r) $ oscillating at frequency $ \omega $ 
satisfies ${\bf  \nabla} ^{2}( n_{1}\partial P / \partial n_{0} ) + {\bf  
\nabla} ( n_{1} \nabla V ) + m \omega ^{2} n_{1} = 0$ with $ \partial 
P / \partial n_{0} = n_{0} \partial \mu / \partial n_{0} $. Of particular 
interest is the 'neutral mode' solution $ n ^{0}_{1}(r)$, corresponding 
to the density fluctuation produced by a small shift $ \delta  \tilde{\mu 
}$ of the overall chemical potential, that is $ n^{0}_{1}(r) = (\partial 
\mu  / \partial n _{0} ) ^{-1} \delta  \tilde{\mu }$. Since the result is 
still an equilibrium situation, this mode corresponds to $ \omega = 0 $ 
(as can be checked directly). Actually this mode does not correspond to 
a physical mode since it does not conserve particle number, whereas 
this number is automatically conserved by all the solutions with non 
zero frequencies, as it can be seen by integrating the starting equations 
over the whole gas volume.

\vspace{4mm}
We make the change $ n_{1}({\bf  r}) = n^{0}_{1}(r)  w(r) Y _{lm}$ 
(i.e. the local fluctuation of the chemical potential is a convenient 
variable). Since $ w(r) = 1 $ corresponds to the neutral mode, we may 
expect a simple equation. Indeed we obtain :
\begin{eqnarray}
rw'' + [ 2 + r L'(r)] w' - [ \frac{l(l+1)}{r} + \frac{ m \omega ^{2} r 
}{ V' (r) }L'(r) ] w = 0
\label{eq1}
\end{eqnarray}
where we have set $ L(r) = \ln (n_{0}(r))$ with $ L'(r) = dL/dr $, and 
$ V'(r) = dV(r)/dr $. This equation for the mode frequencies has a quite 
simple form. In particular, as soon as $ V(r) $ is known, the properties 
of the fluid appear only through the function $ L(r) $, which is itself 
simply related to the equation of state through $ \mu (n_{0}(r)) + V(r) 
= \tilde{\mu }$. Therefore it appears much more convenient to model $ 
L(r) $, rather than $ \mu (n) $. Indeed Eq.(1) lends itself to a very large 
number of specific models with analytical solutions or quasi-analytical 
solutions, as we will see below in the case of the harmonic trap. Before 
proceeding to this case it is also interesting to note that Eq.(1) may be 
written with the form of a one-dimensional Schr\"{o}dinger equation 
(with energy equal to zero) by making the further change $ w(r) = \psi 
(r) / (r \sqrt{ n_{0}(r)}) $. The corresponding potential is found to be $ 
(1/r + m \omega ^{2} /V') L' + (1/4) L'^{2} + (1/2)L'' + l(l+1)/r 
^{2}$ and is again simply related to $ L(r)$. This form is of particular 
interest when one has an explicit analytical solution for an approximate 
model, as we will find below.  One can then easily correct the results by 
a standard first order perturbation calculation. However Eq.(1) is more 
convenient for actual solutions.

\vspace{4mm}
Let us now specialize to the case of the harmonic trap $ V(r) = 
\frac{1}{2} m \Omega ^{2} r^{2}$. It is then convenient to make the 
further change $ w(r) = r ^{l} v(r) $ to get rid of the $ l(l+1)/r $ term. 
This leads to :
\begin{eqnarray}
rv'' + [ 2(l+1) + r L'(r)] v' - ( \nu ^{2} - l ) L'(r) v = 0
\label{eq2}
\end{eqnarray}
where $ \nu ^{2} =  \omega ^{2} /  \Omega ^{2}$. We check on this 
equation that, whatever the equation of state $ \mu (n)$ of the fluid, we 
have as expected the dipole mode ($ l = 1 $) solution corresponding to 
the oscillation in the harmonic trap of the gas as a whole, at frequency $ 
\omega = \Omega $. It corresponds \cite{str} to the solution $ v = 1$ 
with $ \nu ^{2}= 1$. Furthermore $ v = 1$ gives also $ \omega = 
\Omega \sqrt{l} $ whatever $ L(r) $ i.e. independent of the equation of 
state of the fluid (and in particular whether it is a Bose or Fermi gas) 
\cite{ming}. This generalizes for an interacting fluid, at low 
temperature, results obtained by Griffin et al. and Stringari 
\cite{gws,str} for a Bose gas and by Bruun and Clark \cite{brcl} for a 
free Fermi gas. Naturally one can check on the starting equations that 
the corresponding 'surface mode' \cite{gws} fluctuation $ n_{1}(r) = r 
^{l} n^{0}_{1}(r) \propto r ^{l-1} (\partial n_{0} / \partial r) $ is 
indeed solution.

\vspace{4mm}
Let us note first that Eq.(2) is invariant under the change of scale $ r 
\rightarrow Kr $, provided we make the same change of scale for  $ 
L(r)$. This allows us to take in the following the gas radius $R$ as 
unity (consistently with hydrodynamics we work within the Thomas-
Fermi approximation). Next we notice that Eq.(2) is only slightly 
modified by the change of variable $ y = r ^{ \alpha }$ provided again 
that the same change is made for $ L(r)$. This gives \cite{note}  :
\begin{eqnarray}
y \frac{d ^{2}v}{ dy ^{2}}  + ( 1+ \frac{2l+1}{\alpha } + y  
\frac{dL}{ dy})  \frac{dv}{ dy} - \frac{\nu ^{2} - l }{\alpha } 
\frac{dL}{ dy} v = 0
\label{eq3}
\end{eqnarray}
A convenient feature of Eq.(3) is also that the absolute scale in density 
disappears in $L'$ and only $ \bar{n}(r) \equiv n(r)/n(0)$ enters. We 
introduce similarly a normalized local chemical potential $\bar{\mu }(r) 
\equiv \mu (n(r))/\mu (n(0))$ where $ \mu (n(0))$ is simply obtained 
from the gas radius $R$ by $ \mu (n(0)) = \frac{1}{2} m \Omega ^{2} 
R^{2}$, leading to $ \bar{\mu } = 1 - r^{2}$, with $r$ being now in 
units of $R$.

\vspace{4mm}
Looking now for simple situations where we can solve Eq.(3), we 
consider naturally first the case of the non interacting Fermi gas 
\cite{brcl} . This gives $ \bar{\mu } = \bar{n} ^{1/p}$ with $ p = 
3/2$. Similarly we can consider an interacting dilute Bose gas \cite{str} 
where $ \mu = gn$ ($g$ being the coupling constant) leading again to $ 
\bar{\mu } = \bar{n} ^{1/p}$ with now $ p = 1$. These two cases 
imply $ L(r) = p \ln(1- r^{ \alpha })$with $ \alpha = 2 $ in Eq.(3). In 
this way we are lead to consider for any $ \alpha $ and $p$ the model $ 
dL/dy = - p / (1-y) $ for which Eq.(3) becomes explicitely :
\begin{eqnarray}
y (1-y) \frac{d ^{2}v}{ dy ^{2}} + [c - y (p+c)] \frac{dv}{ dy}+ p 
\frac{\nu ^{2}-l }{\alpha } v = 0
\label{eq4}
\end{eqnarray}
with $ c = \!1\!+\! \frac{2l+1}{\alpha }$. The general solution 
\cite{abst} of this equation, giving a non divergent density fluctuation 
for $r = 0$, is the hypergeometric function $ F(a,b;c;y)$, with $ a + b 
= p + c - 1 $ and $ ab = - p \frac{\nu ^{2} - l }{\alpha }$. We have 
furthermore to require that the solutions satisfy the boundary condition 
that the outgoing particle current is zero everywhere on the sphere 
$r=1$. This is not verified \cite{abst} by the general solution, except if 
we require that $a = - n$ where $n$ is a non negative integer in which 
case the solution is a polynomial \cite{str}. This leads to the following 
general result for the normal mode frequencies :
\begin{eqnarray}
\frac{\omega ^{2}}{\Omega ^{2}} = l + \frac{\alpha }{p} \: n \: ( n + 
p + \frac{2l+1}{\alpha })
\label{eq5}
\end{eqnarray}
which agrees naturally with Stringari \cite{str} for $ \alpha = 2$ and $p 
= 1$, and with Bruun and Clark \cite{brcl} for $ \alpha = 2$ and $p = 
3/2$. One may naturally wonder about the interest of these results for 
other values of our parameters $ \alpha $ and $p$. These cases 
correspond to the density distribution $ \bar{n}(r) = ( 1 - r ^{ \alpha }) 
^{p}$ and the equation of state $ \bar{\mu } = 1 - ( 1 - \bar{n} ^{1/p}) 
^{2/ \alpha }$. Our point is that these corresponding models can be 
used to represent closely the equation of state $ \mu (n)$ for a general 
fluid (with a given maximum density $n(0)$). We will show explicitely 
below that the flexibility offered by the two parameters $ \alpha $ and 
$p$ makes it a very convenient and efficient procedure. However at 
first glance these general models do not seem very physical since, 
although their density is properly vanishing at the gas radius, they give 
near this border $ \bar{\mu } \approx \bar{n} ^{1/p}$ whereas one 
should get the dilute gas behaviour $p=1$ (bosons) or $p=3/2$ 
(fermions). However just because the gas is dilute near $ r = 1 $, we do 
not expect this part of the gas to play a significant role. Similarly these 
models give $  \bar{n}  \approx 1 - p r ^{\alpha}$ for small $r$ 
whereas one expects only the case $ \alpha = 2$ to occur for a regular 
equation of state. But if despite of these shortcomings  $  \bar{n }_{0} 
(r) $ is closely approximated over the whole range, one may expect this 
modeling to be already quite reasonable.

\vspace{4mm}
Nevertheless it is clearly of interest to consider more complicated 
models which could display proper behaviour near the center and the 
border of the cloud. Although we have not obtained such models with 
completely analytical solutions, we have found a large class of models 
with quasi-analytical solutions which are in practice not different from 
fully analytical solutions. These are the models $ dL/dy = - \sum 
_{k=0} ^{K}p _{k}y ^{k} / (1-y) $ (where we could take $ \alpha = 
2$ and $ p \equiv \sum _{k=0} ^{K}p _{k} =1$ or $ 3/2$ in order to 
have the proper center and border behaviour). To be simple and specific 
let us take the case $K = 1$, giving  $ - dL/dy = ( p _{0} + p _{1} y ) / 
(1-y) $. This corresponds to the equation of state $ \bar{n } = \bar{ \mu 
} ^{p} \exp [p _{1} (1- \bar{ \mu } )]$. In this case Eq.(3) becomes :
\begin{eqnarray}
y (1-y) \frac{d ^{2}v}{ dy ^{2}}  +  (q _{2}y ^{2}+ q _{1}y + q 
_{0}) \frac{dv}{ dy} + (r _{1}y + r _{0}) v = 0
\label{eq6}
\end{eqnarray}
with $ q _{2} = - p _{1}, q _{1} = -(1+ \frac{2l+1}{\alpha }+ p 
_{0}), q _{0} = 1+ \frac{2l+1}{\alpha }, r _{1} = p _{1} \frac{\nu 
^{2} - l }{\alpha }$ and $r _{0} = p _{0} \frac{\nu ^{2} - l }{\alpha 
}$. When we look for a series expansion of the solution $ v =  \sum 
_{n=0} ^{ \infty }a _{n}y ^{n}$, we find the following recursion 
relation (with $ a _{-1}=0$) : $
[(n+1) (n+q _{0})] a _{n+1}+[-n(n-1)+n q _{1}+r _{0}] a _{n}+[(n-
1)q _{2}+r _{1}] a _{n-1} = 0 $ which does not allow in general for a 
polynomial solution. For large $n$ this relation becomes asymptotically 
$  a _{n+1} - a _{n} =  - (q _{2}/n) a _{n-1}  $. This leads to the 
standard behaviour $ a _{n+1} \approx a _{n}$ giving a convergence 
radius equal to 1. This is the same situation as for the hypergeometric 
function in Eq.(4) (which corresponds to $ q _{2} = r _{1} = 0$) and 
this leads in the same way to a singular behaviour for $y=1$ which 
does not agree with the boundary condition. On the other hand the 
above asymptotic relation may also have solutions $ a _{n+1} \ll a 
_{n}$ implying $ a _{n} \approx (q _{2}/n) a _{n-1}  $ which gives $ 
a _{n} \sim 1/n! $. This very rapidly convergent series has an infinite 
convergence radius and no singularity for $y=1$. It corresponds to the 
physically acceptable solutions. Since we have only to deal with the 
range $ y \in [0,1] $ this solution is a quasi-polynomial since the higher 
order terms in the series are very rapidly negligible. This is quite 
analogous to the polynomial solution of the hypergeometric differential 
equation. Naturally these solutions arise only for special values of our 
parameters, which gives finally the mode frequencies. In practice these 
parameters are found very easily in the following way. We solve 
iteratively the recursion relation for $ a _{n}$ with $ 0 \leq n \leq N $ 
and we require $ a _{N+1} = 0 $ (as if we had a polynomial solution). 
Since $ r _{0}$ and $ r_{1}$ are linear in $ \nu ^{2}-l $, this is 
equivalent to find the roots of an equation of order $N$ for $ \nu ^{2}-l 
$. We then increase the value of $N = 1,2,...$ until the roots have 
converged. For the lowest root this is usually a very fast convergence, 
so one could obtain approximate analytical expressions. But the 
numerics is so easy that this seems unnecessary. Naturally this 
procedure gives immediately Eq.(5) for the hypergeometric case. All 
this analysis and procedure can be extended to the case of $ K > 1 $, 
which gives $ a _{n}\sim (1/n!) ^{1/K}$ for the convergence of the 
series for the physical solution.

\vspace{4mm}
As an example we turn now to a specific case and consider the case of 
two equal populations of fermions in different hyperfine states. This 
may be the case of $^{6}$Li or $^{40}$K near a Feshbach resonance 
\cite{bohn}. We assume an attractive interaction between unlike atoms 
with an interaction $g$, related to the (negative) diffusion length by $ g 
= 4 \pi \hbar ^{2}a / m $, and we take the Hartree approximation to 
describe this system. For total atomic density $n$, the chemical 
potential is given by $ \mu (n) = \hbar ^{2} k _{F}^{2}/2m - | g | n/2 $ 
with $ 3 \pi ^{2} n = k _{F}^{3}$. More generally we could deal with 
situations where $ \mu (n)$ is a polynomial in $u \equiv k _{F}$. In an 
isotropic harmonic trap the equilibrium density satisfies then $ r ^{2} 
\equiv y = P(u)$ where $P(u)$ is a polynomial. In order to solve 
directly this case it is more convenient to rewrite Eq.(2) (taking $ \alpha 
= 2$) with the variable $u$. This leads to:
\begin{eqnarray}
P\!P' v''\!\!+\! [(l\!+\!\frac{3}{2}) P ^{'2}\!\!+\!\frac{3}{u} P\!P'\!\!-
\!\!P\!P''] v'\!-\! \frac{3 (\nu ^{2}\!\!-\!l ) }{2 } \frac{ P ^{'2}}{u}v 
\!=\!0
\label{eq7}
\end{eqnarray}
with $ P'= dP/du$ and $ P'' = d ^{2}P/du ^{2} $. Since this equation 
is homogeneous in $u$ we can rescale this variable and have it varying 
between $0$ and $ 1$. Similarly the homogeneity in $P$ allows to 
conveniently rescale it. The free fermion case corresponds to $ P(u) = 1 
- u^{2}$. After taking $ u^{2}$ as a variable one obtains the 
hypergeometric differential equation which leads again to the free 
fermion result \cite{brcl} $\frac{3}{4}(\nu ^{2}-l) =n ( n + l + 2 ) $. 
For the Hartree approximation we can write $ P(u) = 1 - u^{2} -
\frac{2}{3} \lambda (1 - u^{3})$ with the coupling constant $ \lambda 
= 2 k _{F}(0) | a | / \pi $ and $ k _{F}(0) $ the equilibrium Fermi 
wavevector at the center $r = 0$. This coupling constant goes from 
$0$, for the very dilute regime, to $1$ when we reach at the center the 
instability where the gas is going to collapse under the overall attractive 
atomic interaction. In this case we can not solve Eq.(7) by quasi-
polynomials as Eq.(6) because this equation has five regular singular 
points, instead of only two for Eq.(6). 

\vspace{4mm}
We have solved Eq.(7) numerically, as a function of $ \lambda $, for 
the first three monopole mode frequencies ($l=0$). The results are 
given in Fig.1 . One sees that the frequencies decrease for increasing 
attractive interaction. This was to be expected since the gas gets more 
compressible when one approaches the instability. However, in contrast 
with what one might anticipate, we do not find the lowest mode 
frequency going to zero at the instability. This can be understood 
because the instability density is reached only at the center, and the rest 
of the gas still provides a restoring force accounting for the nonzero 
frequency. Actually this limit can not be reached experimentally since 
the modes correspond to an infinitesimal density oscillation. Any finite 
oscillation will induce nonlinear effects and produce a collapse of the 
gas. So one should experimentally work with ever smaller oscillation 
when one goes near the instability.  We believe also that including 
quantum effects (the hydrodynamic description is not correct at the scale 
of the Fermi wavelength) will, so to speak, smear the region where the 
instability density is reached and lead to a zero frequency mode very 
near the instability (preliminary calculations support this view).

\vspace{4mm}
It is now of interest  to consider an approximate solution of this same 
problem with the modeling we have discussed above. With the Hartree 
approximation the relation between the reduced chemical potential $ 
\bar{ \mu }$ and the reduced density $ \bar{n}$ is $ \bar{ \mu } = (3  
\bar{n} ^{2/3} - 2 \lambda  \bar{n})/( 3 - 2 \lambda ) $. For each value 
of $ \lambda $ we approximate  $ \bar{ \mu }(\bar{n})$ by $ \bar{\mu 
} = 1 - ( 1 - \bar{n} ^{1/p}) ^{2/ \alpha }$, where we obtain the 
parameters $p$ and $ \alpha $ by a least square fit. Then the mode 
frequencies are given by $ \nu ^{2} = \frac{\alpha }{p} n ( n + p + 
\frac{1}{\alpha }) $ for $n = 1,2,3$. The results are given in Fig.1 and 
they are surprisingly close to our exact results from numerical 
integration, although our modeling does not give in general a good 
description of $ \bar{ \mu }(\bar{n})$ neither for $ \bar{n}\approx 0$ 
nor for $ \bar{n}\approx 1$, as already mentioned. The real interest of 
this approximate treatment is that it is very easily inverted and allows to 
analyze readily experimental data. If we have, for a given $ \lambda $, 
the frequencies of the first two modes, we can obtain the values of $p$ 
and $ \alpha $ from the formula for the frequencies (the values for the 
higher modes give a check on this modeling). Hence we have $ \bar{ 
\mu }(\bar{n})$ from our model. By varying the coupling constant we 
can obtain the equation of state $ \mu (n)$ it originates from. In our 
case we could, from the result of Fig.1, recover the Hartree expression 
for $ \mu (n)$ to a very good approximation. This method is much 
more flexible than trying to fit the data with a specific equation of state. 
Naturally it is not clear that this modeling will always work so well and 
we can make use of the richer models with quasi-polynomial solutions 
we have already discussed, as we will see below.
\begin{figure}
\centering
\vbox to 70mm{\hspace{1mm} \epsfysize=7cm \epsfbox{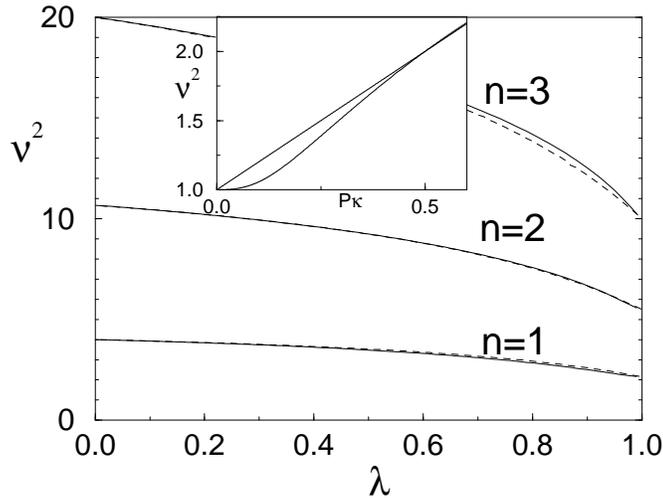} }
\caption{Reduced mode frequency $ \nu ^{2}=( \omega / \Omega ) 
^{2}$ for a Fermi gas within the Hartree approximation as a function of 
the coupling constant $ \lambda $. Full line: exact result from the 
numerical solution of Eq.(7). Dashed line: approximate analytical 
solution. Insert: lowest mode frequency at the collapse ($ \lambda =1$) 
for a generalized equation of state (see text). The straight line is $ \nu 
^{2} = 1 + 2 P _{ \kappa }$. $P _{ \kappa }= 0.577$ for the Hartree 
approximation.}
\label{figure1}
\end{figure}
In this aim we investigate the situation when the gas is right at the 
collapse limit. This corresponds to $ \lambda = 1$ in the Hartree 
approximation. However it is not clear that the Hartree approximation 
will be so good, in particular in the vicinity of the collapse and we may 
consider more general equations of state. The critical density at the 
collapse $ n _{c}$ satisfies $ \partial \mu / \partial n _{c}= 0$. Let us 
assume that $ \mu (n)$ is regular in the vicinity of $ n _{c}$, so we can 
write the expansion $ \mu (n) = \mu  _{c} - \kappa (n - n _{c}) 
^{2}/2$. At the collapse we have $ n(r=0) = n _{c}$ and $  \mu  _{c} 
= \frac{1}{2} m \Omega ^{2} R_{c}^{2}$ where $ R_{c}$ is the gas 
radius at the collapse. In the vicinity of the center $\mu (n) +  
\frac{1}{2} m \Omega ^{2} r^{2}= \mu  _{c}$ gives $ 1 -  \bar{n}(r) 
= (n _{c} - n )/ n _{c} = ( r / R_{c}) ( R_{c}/ L _{ \kappa }) $, where 
we have introduced the length scale $ L _{ \kappa } = (n _{c}/ \Omega 
) ( \kappa /m) ^{1/2}$. This behaviour can be described by our model $ 
\bar{n}(r) = ( 1 - r ^{ \alpha }) ^{p}$ (with now $r$ in units of $ 
R_{c}$ ) if we take $ \alpha = 1$ and $ p = R_{c}/ L _{ \kappa }$. 
Then from Eq.(5) the lowest frequency monopole mode is given by $ 
\nu ^{2} = 1 + 2 / p $. The Hartree approximation corresponds to $ p = 
\sqrt{3} $. The result $ \nu ^{2} = 1 + 2 /  \sqrt{3} = 2.15 $ is in 
excellent agreement with Fig.1 (actually our least square fit gives $ 
\alpha $ and $ p $ very near 1 and $ \sqrt{3}$). If we had chosen $ p = 
3/2 $ in order to have a better description of near $r = 1$, the result $  
\nu ^{2} = 7/3 $ would not be as good. Now we can improve our 
modeling by taking $  \bar{n}(r) =  ( 1 - r ) ^{3/2} \exp (p _{1}r) $ 
which has quasi-polynomial solutions. We have already taken our two 
parameters $ \alpha = 1$ and $ p = 3/2 $ to have a good description 
near the center and the border. We can still adjust $ p _{1} $ to have a 
better description $ \bar{n}(r) = 1 - r  R_{c}/ L _{ \kappa } $ near the 
center. This gives $ p _{1} = 3/2 - R_{c}/ L _{ \kappa }$. Our results 
for the lowest monopole frequency are reported in the insert of Fig.1 as 
a function of the parameter $ P _{ \kappa }= L _{ \kappa }/ R_{c} = ( 
\kappa n _{c}^{2}/2 \mu _{c}) ^{1/2}$. The results of the two models 
are almost indistinguishable in the vicinity of the Hartree case ($P _{ 
\kappa }= 1/\sqrt{3}$) and go to the same limit $ \nu ^{2} = 1 $ for $P 
_{ \kappa }= 0$ (see below). In between they differ somewhat. It is 
clear that $ \bar{n}(r) = ( 1 - r ) ^{p}$ is not flexible enough to give a 
precise description of $ \bar{n}(r) $, whereas $ \bar{n}(r) =  ( 1 - r ) 
^{3/2} \exp (p _{1}r) $ satisfies all the constraints we have set, so we 
can be fairly confident in the results. We see in Fig.1 that the 
experimental determination of the lowest monopole frequency gives 
direct information on the equation of state near critical density since it 
gives immediately $ P _{ \kappa }= ( \kappa n _{c}^{2}/2 \mu _{c}) 
^{1/2}$. Let us also mention that the monopole modes fluctuations are 
divergent  for $r=0$ in this collapse limit, as one could expect 
physically.

\vspace{4mm}
The limit where the parameter  $ P _{ \kappa }$ goes to zero is 
particularly interesting. It corresponds to the situation where most of the 
gas is concentrated at small $r$ and the rest of the gas is irrelevant. One 
may take $ R_{c} \rightarrow \infty $ and take $ L _{ \kappa }$ as new 
length unit for $r$. Then our model $ \bar{n}(r) =  ( 1 - r ) ^{3/2} \exp 
(p _{1}r) $ becomes $ \bar{n}(r) = \exp(- r)$, corresponding to $ 
\bar{\mu  } = 1 - \ln ^{2}(\bar{n})$. The general result is then easily 
obtained from Eq.(2) : since we have merely $ L'(r) = -1 $, Eq.(2) 
reduces to the equation for Laguerre polynomials, and the spectrum for 
all the modes is degenerate and given by $ \nu ^{2} = N $ where $N$ 
is a positive integer. Actually the corresponding Schr\"{o}dinger form 
of the equation, mentionned above, reduces to the one for the hydrogen 
atom. Similarly in our model $ \bar{n}(r) = ( 1 - r ^{ \alpha }) ^{p}$, 
we have $ p \rightarrow \infty $ in this limit. When we take $ L _{ 
\kappa }$ as new length unit for $r$, this model becomes again $ 
\bar{n}(r) = \exp(- r)$ with the same result for the spectrum. Indeed we 
check on Eq.(5) that, in the limit $ p \rightarrow \infty $, the spectrum 
is given by $ \nu ^{2} = l + n \equiv N $ and is degenerate.

\vspace{4mm}
We are very grateful to M. Brachet, Y. Castin, C. Cohen-Tannoudji, J. 
Dalibard and C. Salomon for very stimulating discussions.

\vspace{4mm}
* Laboratoire associ\'e au Centre National de la Recherche Scientifique 
et aux Universit\'es Paris 6 et Paris 7.


\begin{references}
\bibitem{note} Another possible change of variable is  $ y = \ln r $ 
which leads to a very simple equation. It allows also to solve 
analytically some simple models, but they seem less relevant physically 
for our purposes.
\end{references}
\end{document}